\documentclass[11pt]{article}
\usepackage[top=2cm,bottom=3cm,left=2.5cm,right=2.5cm,marginparwidth=1.75cm]{geometry}
\usepackage{graphicx}
\usepackage{amssymb}
\usepackage{amsmath,bm,color}
\usepackage{multicol}
\usepackage{appendix}
\usepackage{graphicx,amsmath,amssymb,bm,epstopdf,epsfig,psfrag,rotate}
\usepackage{mathrsfs,amssymb,mathtools}
\usepackage{cite}
\usepackage{natbib}
\usepackage{float}

\newcommand\Ha{\mbox{\rm{Ha}}}  % Nusselt number
\newcommand\Nu{\mbox{\rm{Nu}}}  % Nusselt number
\newcommand\Ra{\mbox{\rm{Ra}}}  % Rayleigh number
\newcommand\Pran{\mbox{\rm{Pr}}} % Prandtl number
\newcommand\Rey{\mbox{\rm{Re}}}  % Reynolds number
\newcommand\Pm{\mbox{\rm{Pm}}} % magn. Prandtl number
\newcommand\Rm{\mbox{\rm{Rm}}}  % magn. Reynolds number

\newcommand{\bb}[1]{\boldsymbol{#1}}
\DeclareMathOperator\erf{erf}
\DeclareMathOperator\erfc{erfc}

\usepackage{newtxtext}
\usepackage{newtxmath}

\title{\Large{\textbf{Heat transport model for the transition between scaling regimes in quasistatic and full magnetoconvection} }}
\author{Matthew McCormack\textsuperscript{1}, Andrei Teimurazov\textsuperscript{2,}, Olga Shishkina\textsuperscript{2,}, and Moritz Linkmann\textsuperscript{1,}\footnote{moritz.linkmann@ed.ac.uk}}
\date{\small{{\textsuperscript{1}School of Mathematics and Maxwell Institute for Mathematical Sciences, University of Edinburgh, UK}\newline\textsuperscript{2}{Max Planck Institute for Dynamics and Self-Organization, 37077 Göttingen, Germany} }\newline \newline January 9, 2025}

\begin{document}

\maketitle

\vspace{-20pt}
\begin{abstract}
In magnetoconvection, the flow is governed by the interplay between gravitational buoyancy and the Lorentz force, with one of these forces dominating in different regimes.
In this paper, we develop a model with a single adjustable parameter that accurately captures the smooth transition from a buoyancy-dominated regime to one dominated by the Lorentz force. A perturbative extension of the model accounts for distinct transition features that occur at high Prandtl numbers.
We validate the model for magnetoconvection in both the quasistatic regime and at finite magnetic Reynolds numbers using data from direct numerical simulations and existing experimental data sets.
The model contains a natural extension to rotating convection and offers a potential generalisation to rotating magnetoconvection.

\end{abstract}

\vspace{5pt}
\section{Introduction}
\label{introduction}

Understanding convection is a vital building block to our knowledge of fluid mechanics, with fluid flows driven by buoyancy forces being ubiquitous in nature. However, in many such systems, additional forces can greatly affect the resulting flow.  
An example of such an effect occurs in geophysical and astrophysical systems where the flow is highly influenced by the rotation of the body of interest or by the presence of magnetic fields.  
The effect of these forces can also be relevant in a wide variety of industrial processes, in particular, the effect of an imposed magnetic field on liquid metal cooling systems used for cooling nuclear fusion reactors.
In these examples, the effect of an additional force can have a significant effect on the morphology, dynamics, and global transport properties of the flow.
Here, by example of the Lorentz force due to a vertically applied uniform magnetic field, we will investigate the effect of planar forces, which act transversally to the direction of gravity. In this scenario, the additional force acts to suppress convection against buoyancy and tends to align flow structures 
in the direction of the magnetic field.  
In extreme cases of strong applied magnetic field strengths, the force can entirely suppress convection, holding the flow in a motionless conducting state.  
The relative importance of these forces is encoded in dimensionless parameters, which typically measure the force compared to viscous forces in the flow.  
These common dimensionless parameters are the Rayleigh number ($\Ra$) and Hartmann number ($\Ha$), which describe the relative importance of buoyancy and magnetic field strength, respectively.

A central question in the study of convective flows, and perhaps one of the most studied over the past century, is to characterise the heat transport in the flow, given by the Nusselt number ($\Nu$), which is made dimensionless by normalising by the conductive heat transport, in terms of the dimensionless control parameters. The Nusselt number is a key observable of interest which is relatively easy to measure experimentally and is intimately linked to flow's underlying dynamics and formally connected in mean to the turbulent dissipation.  
In Rayleigh--B\'enard convection (RBC) without a constraining force due to rotation or a magnetic field, the Nusselt number is observed to have a power law dependence on the system parameters.  
More precisely, for very large $\Ra$ we have
\begin{equation}
    \Nu \sim \Ra^\gamma,
    \label{eq:intro-r-gamma}
\end{equation}
where $\sim$ denotes equality up to a constant multiplicative prefactor.  
This functional form was originally used as a good fit to data in early experiments of Rayleigh--B\'enard convection but is more formally justified by mathematically rigorous upper bounds on the heat transport.  
Thus, one may consider equation \eqref{eq:intro-r-gamma} as the leading order contribution to $\Nu$ as $\Ra$ becomes very large.  
The remaining question is then to quantify the exponent $\gamma$, the understanding of which at high $\Ra$ is of utmost importance in the context of extrapolating to astrophysical regimes where the value of $\Ra$ is much larger than is currently achievable by experiments or simulations.  

For the classical regime of Rayleigh--B\'enard convection by moderate Rayleigh numbers ($\Ra\lesssim10^{11}$), the Grossmann and Lohse theory \cite{Grossmann2000, Grossmann2001, Grossmann2004, Stevens2013, Ahlers2009} is able to predict accurately the scaling relations for the Nusselt numbers and Reynolds numbers in different regions of the $\Ra-\Pr$ parameter space motivated by the relative importance of the kinetic and thermal dissipation rates in the boundary layers and the bulk region of the flow.
Starting at sufficiently high Rayleigh number, the scaling laws do not change anymore, and the flow approaches the so-called ultimate regime, with an increased effective scaling exponent in the $\Nu$ versus $\Ra$ relation in the ultimate regime compared to the classical regime.
This exponent depends on the Prandtl number and varies from 1/3 to 1/2, subject to some logarithmic corrections \cite{Lohse2023, Lohse2024, Shishkina2024}.

However, it is less clear how the heat transport responds when an additional force is applied to the system, in particular, the Lorentz force.
The heat transport scaling exponent varies strongly with increased constraining force (increased $\Ha$) but also varies locally depending on the value of $\Ra$.  This provides a great challenge. 

In this paper, we derive a model with a single free parameter that fully describes the smooth transition from the buoyancy-dominated regime to the Lorentz-force-dominated regime. 
The model is an extension of our previously proposed parameter-free model \cite{Teimurazov2024} that allowed us to calculate the scaling exponents in the Nusselt number scaling in the Lorentz-force-dominated regime from the scaling exponent in the buoyancy-dominated regime and vice versa.
The model presented here allows us to quantitatively describe the behaviour of the heat transport in the transitional regime between these two regimes.
An additional parameter may be introduced perturbatively in the model to allow us to account for characteristic overshooting effects in the transitional region, specific to high Prandtl numbers.
When this parameter is set to zero, the model gives a smooth and monotonic behaviour of the scaling exponent as a function of the dimensionless control parameters.
We validate the model against data from direct numerical simulations (DNSs) \cite{Teimurazov2024} of liquid metal convection in the quasistatic regime, using the computational code {\sc goldfish} \cite{Shishkina2015, Reiter2020, Reiter2022, Teimurazov2024}, and data from the literature obtained by numerical simulations \cite{Lim2019, Akhmedagaev2020, Xu2023} and laboratory experiments \cite{Cioni2000, King2015, Zuerner2020, Xu2023}.  
Furthermore, we demonstrate its applicability to full magnetoconvection (MC). 
Finally, we briefly discuss the potential applications of the model to rotating convection and rotating magnetoconvection.  

\section{Equations of motion and dimensionless parameters}
We consider three-dimensional convection between two infinitely wide and long plates separated by a height $H$, driven by an imposed temperature difference between the bottom and top plates $\delta T = T_{\rm bottom} - T_{\rm top}>0$, and subjected to a (quasi)planar force $\bb{f}_\perp$ which may either be due to rotating or magnetic effects, and, depending on system parameters, may be largely or completely restricted to the cross-plane direction. The equations of motion under the Oberbeck-Boussinesq approximation are
\begin{subequations}
\label{eq:governing}
\begin{alignat}{1}
    \partial_t\boldsymbol{u} + \boldsymbol{u}\cdot\nabla\boldsymbol{u}+\nabla (p/\rho_0) &= \nu \Delta\boldsymbol{u} + \alpha g(T-T_0)\boldsymbol{e}_z + \boldsymbol{f}_\perp, \\
    \partial_t T + \boldsymbol{u}\cdot\nabla T &= \kappa\Delta T, \\
    \nabla \cdot \boldsymbol{u} &= 0,
\end{alignat}    
\end{subequations}
where $\bb{u}(\bb{x},t)$ is the velocity, $T(\bb{x},t)$ the temperature, $p(\bb{x},t)$ the kinematic (or reduced) pressure, $\rho_0$ the reference fluid density, $\nu$ the kinematic viscosity, $\alpha$ the thermal expansion coefficient, $g$ the acceleration due to gravity, $T_0$ the reference temperature, and $\kappa$ is the thermal diffusivity.  
We consider no-slip boundary conditions $(\bb{u}=0)$ on the top and bottom boundaries.

The governing dimensionless parameters of the system are the Rayleigh number $\Ra$ and the Prandtl number $\Pran$
\begin{equation}
    \Ra = \frac{\alpha g \,\delta T H^3}{\kappa \nu}, \qquad \Pr = \frac{\nu}{\kappa},
\end{equation}
which control the dimensionless temperature difference applied across our fluid layer (i.e. the thermal driving), and the ratio of momentum diffusion to heat diffusion in the system respectively.

\subsection{Magnetoconvection with a vertical background magnetic field}
If we consider an electrically conducting fluid subjected to a uniform vertically orientated background magnetic field $B_0$, the force $\bb{f}_\perp$ is given by the Lorentz force
\begin{equation}
    \bb{f}_\perp = \frac{1}{\rho_0}(\bb{J}\times\bb{B}),
\end{equation}
where $\bb{B}(\bb{x},t) = B_0\bb{e}_z + \bb{b}(\bb{x},t)$ is the total magnetic field comprised of the background ($B_0$) and fluctuating ($\bb{b}$) magnetic field, where $\bb{e}_z$ is the vertically orientated unit vector, and $\bb{J}$ is the current density which is calculated through Ohm's law
\begin{equation}
    \bb{J} = \sigma(\bb{E} + \bb{u}\times\bb{B}) = \frac{1}{\mu}(\nabla\times\bb{B}),
\end{equation}
where $\sigma$ is the electrical conductivity, $\bb{E}$ the electric field, and $\mu$ the magnetic permeability. 

Additionally, Maxwell's equations can be combined into a single equation known as the induction equation
\begin{equation}
    \partial_t \bb{B} + \bb{u}\cdot\nabla\bb{B} = \bb{B}\cdot\nabla\bb{u} + \eta\Delta\bb{B},
    \label{eq:induction}
\end{equation}
where $\eta = (\mu\sigma)^{-1}$ is the magnetic diffusivity.  The magnetic field must additionally be solenoidal $\nabla \cdot \bb{B}=0$ due to Gauss's law. 

In this case, two new control parameters are added to the problem, the Hartmann number $\Ha$ and the magnetic Prandtl number $\rm{Pm}$
\begin{equation}
    \Ha = B_0 H \sqrt{\frac{\sigma}{\rho_0 \nu}}, \qquad \rm{Pm} = \frac{\nu}{\eta},
\end{equation}
which control the relative strength of the applied magnetic field, and the ratio of momentum diffusion to magnetic diffusion respectively. The Hartmann number is also equivalent to the Chandrasekhar number $Q = \Ha^2$.

A common approximation to this system, known as the quasistatic approximation, is realised in the case where $\rm{Pm}\ll1$, provided that the magnetic Reynolds number $\rm{Rm} = U\ell/\eta\ll1$, where $U$ and $\ell$ are characteristic velocity and length scales. In this case, any fluctuating part to the induced magnetic field is negligible and $\bb{B} \approx B_o\bb{e}_z$. The result is that the induction equation Eq.~\eqref{eq:induction} does not need to be considered in the formulation of the problem. In this case, $\bb{E} = -\nabla\phi$ , where  $\phi(\bb{x},t)$ is the electric field potential. The divergence-free property of the current density is
then used to calculate the electric field potential
through the following Poisson equation,
\begin{equation}
    \Delta \phi = \nabla \cdot (\bb{u} \times \bb{B}).
\end{equation}
The quasistatic approximation is an excellent approximation in liquid metal convection and is thus relevant to most experimental studies and industrial applications of magnetoconvection.

\section{Formulation of the heat transport model}
\label{sec:heat_transp_model}
In the following section, we introduce a model for the heat transport in the system, which is given in dimensionless form by the Nusselt number
\begin{equation}
    \Nu = \frac{\langle u_zT\rangle_z - \kappa \partial_z \langle T\rangle_z}{\kappa\, \delta T/H},
\end{equation}
which measures the total heat flux relative to the conductive heat flux, where $\langle\cdot\rangle_z$ denotes the time average taken over cross sections at height $z$. 

\subsection{Connection between buoyancy-dominated \& Lorentz-force-dominated scaling}

When the flow is turbulent with a very weak or no constraining force, the convective heat transport ($\Nu-1$) displays a power-law dependence on the thermal driving $(\Ra)$
\begin{equation}
    \Nu - 1 \sim (\Ra/\Ra_c)^\gamma = a_1 \Ra^\gamma,
    \label{eq:buoy-dom}
\end{equation}
for some exponent $\gamma$.  In this regime, $\Ra_c$ is the critical $\Ra$ for bulk onset in RBC which is a fixed constant for a given container geometry.

However, when the flow is strongly influenced by a constraining force, the heat transport in the system at a given $\Ra$ is heavily influenced, decreasing the total heat transport. For example, when the flow is constrained by a sufficiently large vertically imposed uniform magnetic field with $\Ra>\Ra_c$, the convective heat transport also displays a power-law dependence
\begin{equation}
    \Nu - 1 \sim (\Ra/\Ra_c)^\gamma = a_2 (\Ra/\Ha^2)^\xi,
    \label{eq:mag-dom}
\end{equation}
for some exponent $\xi$, where now, $\Ra_c$ is influenced by the strength of the magnetic field.  Specifically, the onset of bulk convection is delayed by the increased magnetic field with $\Ra_c \sim \Ha^2$ from the linear theory of Chandrasekhar \cite{Chandrasekhar1961}.

Although these scaling laws in the two extreme regimes appear disconnected, a recent model in rotating convection \cite{Ecke2023}, which has been extended for quasistatic magnetoconvection \cite{Teimurazov2024}, suggests that the exponents $\xi$ and $\gamma$ can be connected under the assumption that the transition is controlled by the relative thickness of the viscous and thermal boundary layers which we summarise here for completeness.

We assume that in an intermediate regime, where neither the effect of buoyancy or the Lorentz force are strongly dominant, the two power-laws \eqref{eq:buoy-dom}, \eqref{eq:mag-dom} must overlap, that is
\begin{equation}
    \Ra^\gamma \sim \Nu - 1 \sim (\Ra/\Ha^2)^\xi,
\end{equation}
at an intermediate value of $\Ra$ and $\Ha$ for each given $\Pr$.  We then make two modelling assumptions, the first of which assumes that the transition is primarily controlled by the viscous and thermal boundary layers in the sense that at this intermediate point where the two power laws overlap, the thermal boundary layers $\delta_T$ and Hartmann boundary layers $\delta_\nu$ on the top and bottom plates are related as $\delta_T = \lambda\delta_\nu$ for a constant $\lambda = \lambda(\Pran)$. The second assumption is that both of these boundary layers scale with the control parameters in accordance with well-established laminar theory. In this case, that $\delta_T \propto \Nu^{-1}$ and that $\delta_\nu \propto \Ha^{-1}$.  Both of these assumptions are well established in the rotating case \cite{King2009, King2012, Ecke2023}, and have been seen to hold in all of the currently available data in the magnetic case \cite{Teimurazov2024}. Under these assumptions, one can obtain the following relation between the two scaling exponents
\begin{equation}
    \xi = \frac{\gamma}{1-2\gamma}, \quad \rm{or} \quad \gamma = \frac{\xi}{1+2\xi}.
    \label{eq:xi-gamma-rel}
\end{equation}

This provides a direct relationship between the two regimes which is extremely useful in practice due to the extensive work in the RBC literature for predictions of the exponent $\gamma$, where data from simulations, experiments and validated theoretical predictions are abundant \cite{Grossmann2000, Ahlers2009}. Subsequent predictions of the exponent $\xi$ in the less studied Lorentz-force-dominated regime have been validated against numerical and experimental data at $\Pran = 8$ and $\Pran \approx 0.025$ in \cite{Teimurazov2024}. Prior studies have also validated the equivalent model in the rotating case at two values of $\Pran$ \cite{Ecke2023}.

\subsection{A model for the transition between the two extreme regimes}
\label{sec:transition-coordinates}
\begin{figure}[h]
    \centering
    \includegraphics[width = 0.65\columnwidth]{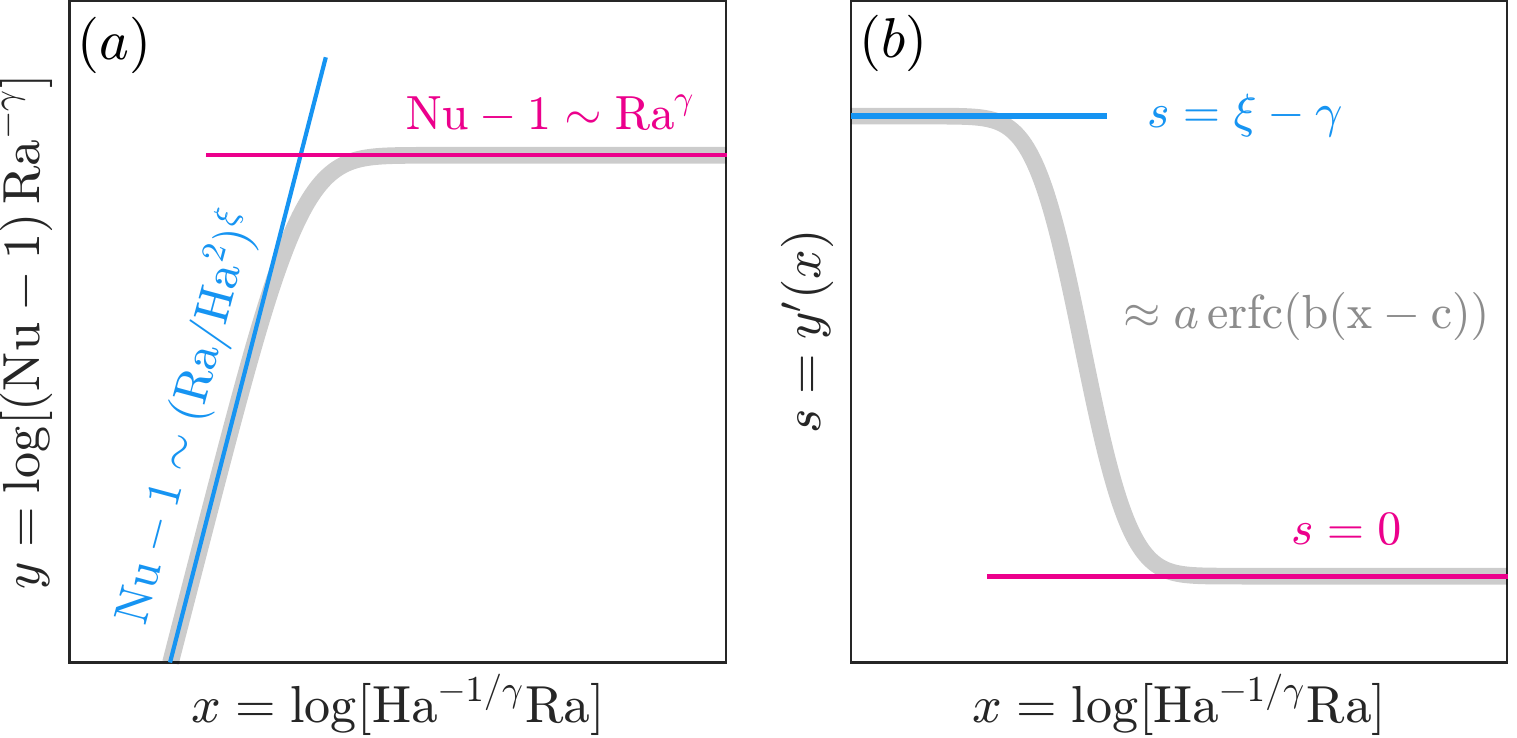}
    \caption{$(a)$ Predicted collapse of the data onto a master curve under the coordinate transform $x=\log[\Ha^{-1/\gamma}\Ra]$ and $y=\log[(\Nu-1) \Ra^{-\gamma}]$ and $(b)$ the slope of the master curve $s(x)$. In both figures, the buoyancy-dominated scaling law (pink) and the  Lorentz-force-dominated scaling law (blue) are shown which control the asymptotic behaviour of the master curve.}      
    \label{fig:collapse-theory}
\end{figure}

It is further shown in Refs.~\cite{Ecke2023, Teimurazov2024}, that the transition between the two extreme regimes can be collapsed onto a master curve through the construction of a scaling law which combines the scaling relations of both regimes
\begin{equation}
    (\Nu-1) \Ra^{-\gamma} \sim [\Ha^{-2\xi/(\xi-\gamma)}\Ra]^s = [\Ha^{-1/\gamma}\Ra]^s,
    \label{eq:transition-scaling}
\end{equation}
which has been expressed only in terms of the exponent $\gamma=\gamma(\xi)$ using Eq.~\eqref{eq:xi-gamma-rel} for convenience.  Notably, this form recovers the buoyancy-dominated scaling law  \eqref{eq:buoy-dom} for $s=0$, and the  Lorentz-force dominated scaling law \eqref{eq:mag-dom} for $s=\xi-\gamma=2\gamma^2/(1-2\gamma)>0$.  It is thus expected that the parameter $s$ controls the transition between the two regimes in the sense that varying $s$ between $0$ and $\xi-\gamma$ would appropriately modulate the scaling of $\Ra$ and $\Ha$ to trace out a master curve. However, independently of $s$, one can test the validity of the constructed scaling relation \eqref{eq:transition-scaling} through the construction of the plot shown in Fig.~\ref{fig:collapse-theory}a below.  In this construction, $s$ now represents the slope of the master curve shown in grey which is sketched for clarity in Fig.~\ref{fig:collapse-theory}b. 

In the present literature, a number of distinct intermediate scaling laws of the form $\Nu \sim \Ra^{\,\beta}$ between the extreme regimes have been observed, with a range of scaling exponents $\beta$ being measured depending on the value of $\Ha$ and the range of $\Ra$ considered. In this view, one may expect the master curve sketched in Fig.~\ref{fig:collapse-theory} to be close to piecewise continuous, with the derivative sharply jumping from one intermediate scaling law to the next. However, a construction of the master curve in practice is suggestive of a higher degree of regularity, with a smooth transition between the two extreme scaling laws being observed \cite{Ecke2023, Teimurazov2024}. This suggests that previously identified intermediate scaling laws should be viewed as locally tangent approximations to a smooth underlying surface in parameter space at least with respect to this choice of coordinates.  

The coordinate transformation motivated by Eq.~\eqref{eq:transition-scaling} additionally has a geometric interpretation. If one considers the three-dimensional parameter space $(\Ra,\Ha,\Nu-1)$ in logarithmic scale, then the coordinate transformation can be decomposed into a rotation, stretching, and a subsequent projection onto the two-dimensional space shown in Fig.~\ref{fig:collapse-theory}a. Interestingly, it appears that a collapse of the data can be nearly fully attributed to the rotational part of the transformation. In this sense, the physical argument extracts an intrinsic dimensional reduction of the $\Nu=\Nu(\Ra,\Ha)$ surface by finding near uniformity in a particular direction, allowing for the collapse onto a master curve. Additional details about the geometric interpretation of the coordinate transform are included in \ref{sec:appendix_geo}.

\subsection{Parameterisation of the master curve} \label{sec:parameterisation-master-curve}
Although the current coordinate transformation collapses the data onto a master curve, the current formulation does not parameterise the curve and thus, does not allow for a prediction of heat transport scaling or the Nusselt number itself in the transitional regimes. However, the non-trivial coordinate transformation motivated by the model (i.e. the coordinates in Fig.~\ref{fig:collapse-theory}a) does motivate a class of simple analytic models that capture heat transport scaling globally across the considered parameter space, formed through the use of sigmoid functions. We will show that such models can be constructed with a very small number of free parameters and are able to reproduce Nusselt numbers across parameter space with reasonable accuracy.

We begin by choosing a change of variables motivated by our model, namely $x=\log[\Ha^{-1/\gamma}\Ra]$ and $y=\log[(\Nu-1) \Ra^{-\gamma}]$. We then note that to a first approximation, the slope of the master curve, i.e. $s = s(x)$ has the general form of a sigmoid function. One can reconstruct the master curve $y = y(x)$ by integrating $s(x)$ directly. Here, we choose the slope to be modelled by the complementary error function
\begin{equation}
    \erfc(z) = 1-\erf{(z)} = \frac{2}{\sqrt{\pi}}\int_z^\infty \exp({-t^2})\,dt,
\end{equation}
where $\erf$ denotes the error function.  In the discussion below, it will be useful to note that $\erfc(z\rightarrow\infty) = 0$, and $\erfc(z\rightarrow -\infty) = 2$.

We construct the master curve $y(x)$ through a direct integration as
\begin{align}
    y(x) = \int s(x)\, dx &= \int a \erfc(b(x-c))\,dx \nonumber \\ &= a(x-c)\erfc(b(x-c)) - \frac{a}{b\sqrt{\pi}}\exp{(-b^2(x-c)^2)} + C,
    \label{eq:y_function}
\end{align}
where $a,b,c$ are free parameters and $C$ is the constant of integration.  In the following discussion, we will refer to this as the $\erfc$ model.

Although we have introduced a number of free parameters in this construction, we will now show that nearly all of these constants may already be determined from existing knowledge of the system.  In particular, we see that only a single free parameter $b$ remains free.  Examining the scaling relation in Eq. \ref{eq:transition-scaling}, we first note that the slope $s$ will asymptotically approach $s \rightarrow \xi-\gamma$ as we move to the  Lorentz-force-dominated regime, and will approach $s\rightarrow0$ in the buoyancy-dominated regime, as illustrated in Fig.~\ref{fig:collapse-theory}.
Imposing these values as boundary conditions on our modelled function $s(x)$ will set the constant $a$.  Namely, we consider the limit of $s$ as $x-c \rightarrow -\infty$, using the properties of the error function stated above, and conclude that $a = (\xi-\gamma)/2 = \gamma^2/(1-2\gamma)$.  We further confirm that $s\rightarrow0$ as $x-c\rightarrow\infty$.

We now turn our attention to imposing boundary conditions on the master curve itself $y(x)$.  Since $y=\log[(\Nu-1) \Ra^{-\gamma}]$, we know that this expression should recover the prefactor $a_1$ (see Eq.~\ref{eq:buoy-dom}) in the buoyancy-dominated regime.  Thus, we must impose in our model that $y(x)\rightarrow \log(a_1)$ as $x-c\rightarrow\infty$, finding that the constant of integration $C = \log(a_1)$.  We now must impose the prefactor in the  Lorentz-force-dominated regime.  When $x-c\ll1$, we observe that $y(x)=(\xi-\gamma)(x-c) + \log(a_1)$.  On substitution of the original variables and imposing that this expression approaches the Lorentz-force-dominated scaling law, the constant $c$ can be obtained in terms of the buoyancy-dominated and  Lorentz-force-dominated prefactors and the corresponding scaling exponents as
\begin{equation}
    c = \frac{\log(a_2/a_1)}{\gamma-\xi} = \frac{(2\gamma-1)\log(a_2/a_1)}{2\gamma^2}.
    \label{eq:c}
\end{equation} 

If we assume that the exponents and prefactors of eqs.~\eqref{eq:buoy-dom} and \eqref{eq:mag-dom} are known from existing theory or data, we can construct this analytic model $\Nu = \Nu(\Ra,\Ha)$ with only a single free parameter $b$.  
Thus in principle, the transition region can be fit using only a single data point in this regime using Eq.~\ref{eq:y_function}, once the other constants have been calculated.  Estimates of $b$ may be obtained with higher accuracy using knowledge of multiple data points using standard fitting methods.
Further, the dependence on $\Pran$ is most heavily contained in the buoyancy-dominated prefactor $a_1$ and the corresponding scaling exponent $\gamma$, which are typically well known.  

Inverting the coordinate transformation, yields an explicit approximation for $\Nu=\Nu(\Ra,\Ha)$
\begin{multline}
    \log(\Nu-1) = \log(a_1\Ra^\gamma) 
    + \frac{\gamma^2}{1-2\gamma}\big[\log(\Ha^{-1/\gamma}\Ra) - c\big]\erfc\big[b(\log(\Ha^{-1/\gamma}\Ra)-c)\big] \\- \frac{\gamma^2}{b(1-2\gamma)\sqrt{\pi}}\exp\big[-b^2(\log(\Ha^{-1/\gamma}\Ra)-c)^2\big],
    \label{eq:nu(ra,ha)}
\end{multline}
where $c$ is given through Eq.~\eqref{eq:c}.  We refer to this as the $\erfc$ model.

It is worth noting that although we have chosen to model the slope of the master curve using the error function, this choice is not unique, and indeed other sigmoid functions could have been used, meaning our approach defines a family of models.  However, the choice of the error function has been made here as it allows for the recovery of the free parameters in closed form, a property that is not true of other choices of sigmoid functions such as a hyperbolic tangent profile for example.  We further note that the method here may be conveniently generalised by viewing the slope of the master curve as the solution to a simple nonlinear boundary value problem which we discuss in \ref{sec:choice_of_sigmoid}.

\subsection{High Pr correction}
\begin{figure}[h]
    \centering
    \includegraphics[width = 0.5\columnwidth]{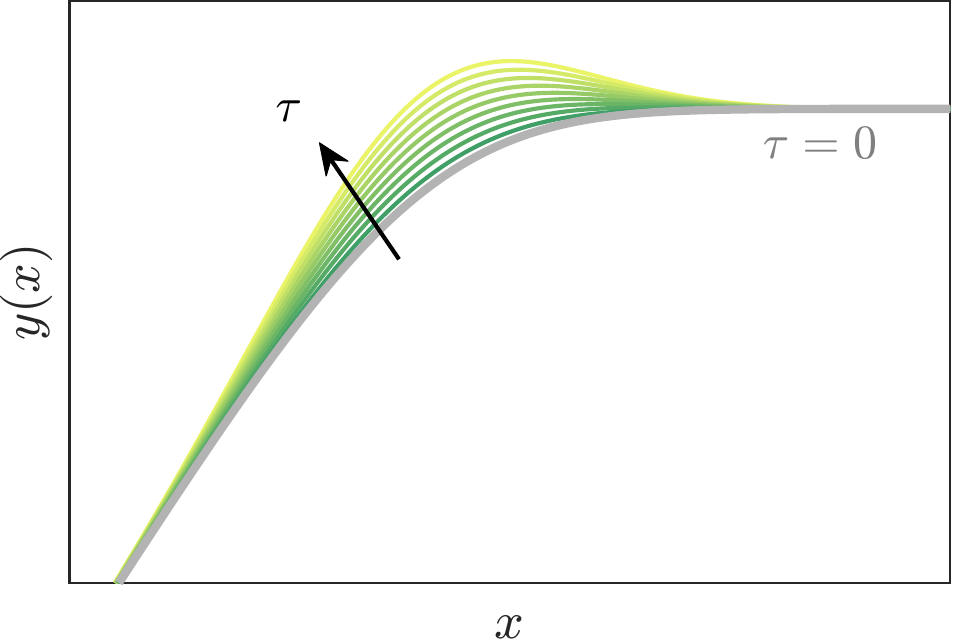}
    \caption{Effect of the correction parameter $\tau$ on the master curve.  $\tau=0$ curve (grey) represents the unperturbed case. Increasing $\tau$ increases the level of overshoot.}  %\textcolor{red}{PLACEHOLDER}}
    \label{fig:alpha_sketch}
\end{figure}

Although most examples of magnetoconvection occur at low Prandtl numbers, some flows of electrolytes or molten salts at higher $\Pran$ may still be influenced by the presence of magnetic fields \cite{Huboda2018,Li2024}, and thus it is relevant to access the model's validity at high $\Pran$.  Although limited data exists in these regimes, a large dataset is available for $\Pran = 8$ \cite{Lim2019}.  This dataset shows a small characteristic overshoot in the master curve which is observed near the transition to the buoyancy-dominated scaling law, which is also seen in rotating convection where it is much more pronounced and has been linked to Ekman pumping.
A further benefit of using the error function to model the slope of the master curve is that it cleanly allows for a correction to account for this overshoot by introducing only a single extra parameter $\tau$.  Namely, the amplitude of the Gaussian term in the master curve may be perturbed as
\begin{equation}
    y(x) = a(x-c)\erfc(b(x-c)) - (1-\tau) \frac{a}{b\sqrt{\pi}}\exp{(-b^2(x-c)^2)} + \log(a_1).
\end{equation}
In the following sections, we will refer to this as the $\erfc$-$\tau$ model. A schematic illustrating the effect of varying $\tau$ is shown in Fig.~\ref{fig:alpha_sketch}.

\section{Validation of the model}
In this section, we provide a validation of our model against DNS data obtained by us and from the literature for quasistatic magnetoconvection and we demonstrate its applicability at finite magnetic Reynolds number when magnetic-field fluctuations need to be taken into account. 

For the quasistatic case, we generated data by direct numerical simulations using the high-order finite-volume solver {\sc goldfish} \cite{Shishkina2015, Reiter2020, Reiter2022}, which has been further developed for quasistatic magnetoconvection by \citet{Teimurazov2024}. The computational code uses a consistent and conservative scheme as described by \citet{Ni2012}. 
The computational domain is cubic, with solid walls in all directions. The boundary conditions are no-slip for the velocity field, and electrically insulating in all directions. For the temperature, the top and bottom plates are at fixed temperature $T_{\rm bottom} > T_{\rm
top}$, while the side walls are adiabatic. 
In all simulations, the grid resolution meets the necessary criteria to accurately resolve turbulent microscales, both in the bulk and within the viscous (Hartmann) and thermal boundary layers \cite{Shishkina2010}.
For further details on the dataset, see Ref.~\cite{Teimurazov2024}.

Full MC simulations at $\Pran =0.25$ and a magnetic Prandtl number $\Pm = 0.5$ 
have been done using Dedalus v2 \cite{burns2020dedalus}
with an adapted version of the script provided by \cite{cresswell2023force} for a range of Hartmann numbers 
$10 \leqslant \Ha \leqslant 3000$ and Rayleigh numbers $5\times 10^6 \leqslant \Ra \leqslant 10^8$.
The values of $\Pran$ and $\Pm$ have been chosen here to match the existing dataset provided by \cite{cresswell2023force}.
The equations of motion are solved in a three-dimensional rectangular domain with solid top and
bottom walls and periodic extension in the remaining directions. The boundary
conditions at the solid walls are fixed temperature $T_{\rm bottom} > T_{\rm
top}$, no-slip for the velocity field, and magnetically conducting.  Dedalus
uses standard pseudospectral methods, here Fourier expansions in the
homogeneous horizontal directions and Chebyshev-expansions in the wall-normal
direction, with 3/2-dealiasing in all directions. The equations of motion are
stepped forward in time by an implicit-explicit third-order four-step
Runge-Kutta scheme.

To compare the model to data, parameter scans in both $\Ha$ and $\Ra$ are required, which results in considerable 
computational cost for full MC. As the purpose of the comparison with full MC data is to establish its scope rather than validation, 
and because of the aforementioned computational effort that is necessary to do this, we needed to compromise on resolution and run time of our simulations. 
Most simulations have been run using $N_x \times N_y \times N_z = 128 \times 128 \times 200$ grid points. The run times range from  
30-40 free-fall times for low $\Ha$ (10-100), about 50-80 free-fall times for intermediate $\Ha$ (200-1000), and 200-400 free-fall times for the high $\Ha$ cases (2000-3000).
For basic convergence checks, short higher-resolution simulations at $N_x \times N_y \times N_z = 200 \times 200 \times 300$ grid points have been run for isolated cases at high $\Ha$ 
that are particularly relevant for the  Lorentz-force-dominated scaling law.
These high $\Ha$ cases have been run for around 100 free-fall times at this resolution and show good agreement with the longer-time low-resolution runs (not shown).
Typically in these regimes, high resolution is needed to capture the thin Hartmann boundary layers on the top and bottom surfaces of the domain.  Since $\delta_\nu \sim \Ha^{-1}$, the boundary layers will be thinnest at our highest magnetic field strength $\Ha=3000$.  At $\Ha=3000$, there approximately 3 grid points within the boundary layer for the $128 \times 128 \times 200$ grid and approximately 4-5 grid points within the boundary layer for the $200 \times 200 \times 300$ grid.  Thus we have a similar number of grid points to previous quasistatic simulations which obtained converged global $\Nu$ values \cite{Teimurazov2024}.
The simulations exhibit non-negligible fluctuating magnetic field effects in line with those observed in simulations at comparable parameter values, discussed in detail by \citet{cresswell2023force}.  
We note that some differences in the results are observed between the simulations performed here and the three-dimensional simulations performed in \cite{cresswell2023force} at comparable parameter values, which are likely a result of the higher grid resolutions used in the present simulations.
An overview of the performed simulations and results are tabulated in \ref{sec:full_mhd_data}.

\begin{figure}[h]
    \centering
    \includegraphics[width = 0.49\columnwidth]{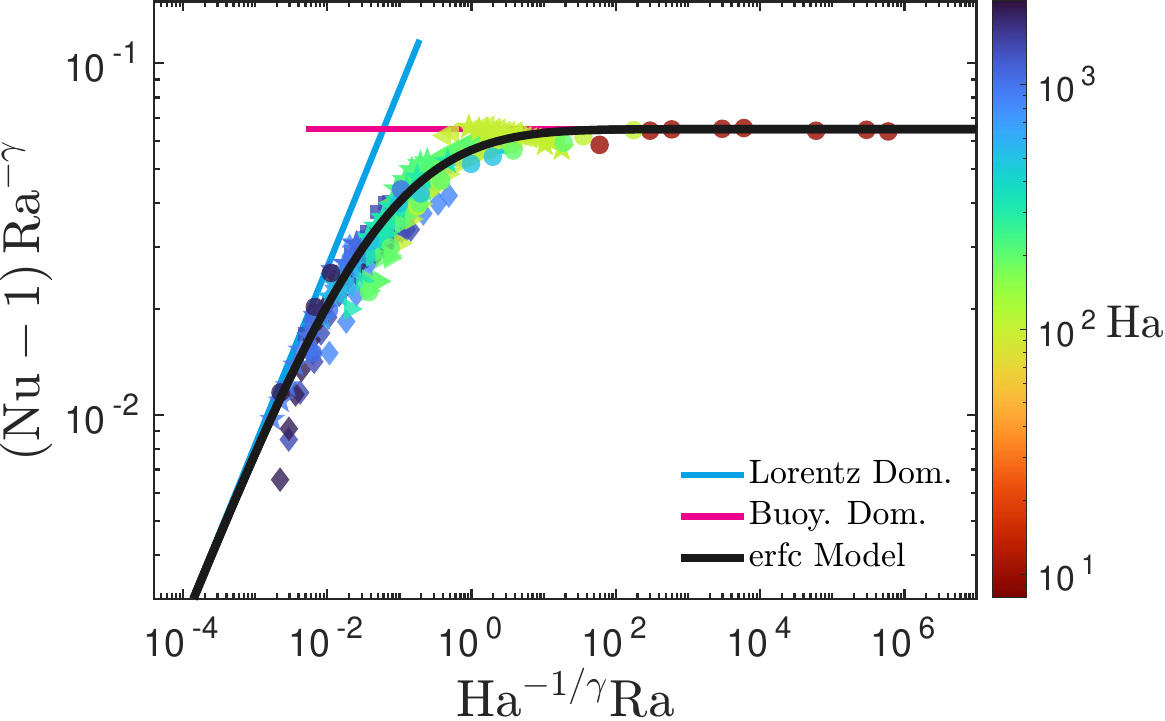}
    \includegraphics[width = 0.49\columnwidth]{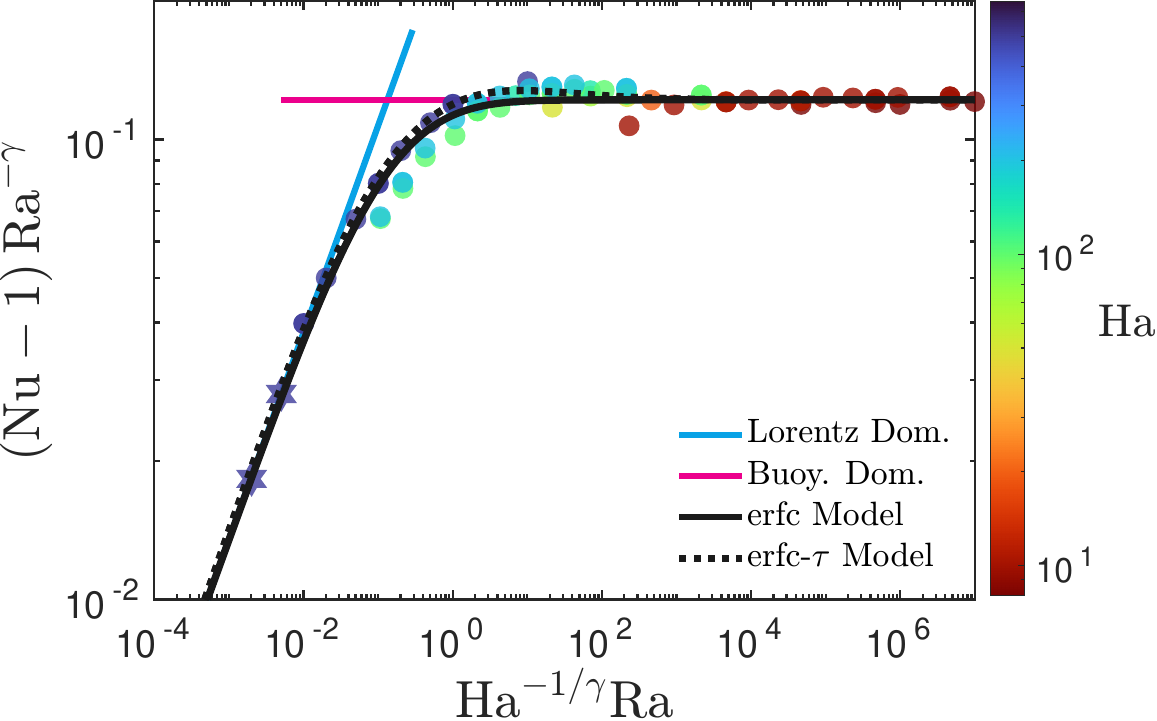}
        \caption{Comparison of the model and datasets for $0.025 \leqslant \Pran \leqslant 0.029$ (left) and $\Pran = 8$ (right). The blue and pink solid straight lines indicate the expected scaling laws for the Lorentz-force (blue) and the buoyancy-dominated (pink) regimes derived in Ref.~\cite{Teimurazov2024}. The erfc-model is indicated by the black solid line, the dashed black line corresponds to the erfc-$\tau$ model. The symbols in the top panel correspond to data from Refs.~\cite{Akhmedagaev2020} (squares), \cite{Cioni2000} (diamonds), \cite{Xu2023} with aspect ratio 1 (left triangles) and aspect ratio 2 (right triangles), \cite{King2015} (pentagrams) and our DNS \cite{Teimurazov2024} (circles). The symbols in the bottom panel correspond to data from Ref.~\cite{Lim2019} (circles) and our DNS \cite{Teimurazov2024} (hexagrams).
    The deviation between the model and data at low $\Ra$ and high $\Ha$ is due to the occurrence of wall modes with very little bulk convection as discussed in Ref.~\cite{Teimurazov2024}.}  
    \label{fig:quasi}
\end{figure}

\subsection{Quasistatic magnetoconvection}
We validate the model against several datasets from the literature obtained from  
DNS \cite{Lim2019, Akhmedagaev2020, Xu2023} and laboratory experiments \cite{Cioni2000, King2015, Zuerner2020, Xu2023}, and our own DNS \cite{McCormack_Teimurazov_Shishkina_Linkmann_2023, Teimurazov2024} at $0.025 \leqslant \Pran \leqslant 0.029$ and $\Pran = 8$, and for $10 \leqslant \Ha \leqslant 2000$. 

We show collapse of the data at $\Pr=0.025$ and $\Pr=8$ and the full range of $\Ha$ in Fig.~\ref{fig:quasi}.  
The solid black line denotes the $\erfc$ model and the dotted black line denotes the $\erfc$-$\tau$ model which we additionally plot for the $\Pr = 8$ data \cite{Lim2019, McCormack_Teimurazov_Shishkina_Linkmann_2023, Teimurazov2024} that shows the small high-$\Pran$ overshoot for the data from Ref.~\cite{Lim2019}.  Blue and pink lines indicate the scaling laws in the Lorentz-force and buoyancy-dominated regimes, (eq. \ref{eq:mag-dom}/\ref{eq:buoy-dom}) respectively \cite{Teimurazov2024}.

\begin{table}[h]
\centering
\begin{tabular}{l c c c c c} 
 \hline
 \quad\:\: Model & $\gamma$ & $a_1$ & $a_2$ & $b$ & $\tau$  \\ 
 \hline
 $\Pr=0.025$ ($\erfc$)    & 0.31 & 0.065 & 0.270 & 0.55 & 0    \\ 
 $\Pr=8$ ($\erfc$)        & 0.30 & 0.122 & 0.302 & 0.80 & 0    \\ 
 $\Pr=8$ ($\erfc$-$\tau$) & 0.30 & 0.122 & 0.302 & 0.50 & 0.45 \\ 
 \hline
\end{tabular}
\caption{Overview of the parameters used in the $\erfc$ and $\erfc$-$\tau$ models for quasistatic MC at the studied values of $\Pr$.  Parameters $\gamma$, $a_1$ and $a_2$ are known from existing literature.  Quantities $b$ and $\tau$ are free parameters of the theory.}
\label{tab:quasistatic_fit}
\end{table}

An overview of the parameter choices used for these models is given in table \ref{tab:quasistatic_fit}.  We emphasise here that the parameters $\gamma$, $a_1$ and $a_2$ are well estimated from existing literature, and thus we only need to fit the parameter $b$ for the $\erfc$ model, or $b$ and $\tau$ for the $\erfc$-$\tau$ model. 
 For the $\erfc$ model, the parameter $b$ can be obtained from a single data point in the transitional regime using Eq.~\ref{eq:nu(ra,ha)}, or may be averaged over a number of data points if additional data is available.  Here, we have averaged the value of $b$ over a number of data points in the transitional regime, finding that the value of $b$ is reasonably insensitive to the particular data point considered, and that a good fit to the data is obtained easily.  A similar approach has been taken for the $\erfc$-$\tau$ model, although better results were typically obtained by fitting the overshoot parameter $\tau$ first to the level of overshoot with an estimate of $b$, and then subsequently fine-tuning $b$ using the data.
\begin{table}[h]
\centering
\begin{tabular}{c c c c } 
 \hline
 $\Ha$ & $\Ra$ & $\beta_{pow}$ & $\beta_{\erfc}$  \\ 
 \hline
 450 & 10$^7$-10$^9$ & 0.420$\pm$0.008 & 0.432 \\ 
 650 & 10$^7$-10$^9$ & 0.476$\pm$0.007 & 0.482 \\ 
 850 & 10$^7$-10$^9$ & 0.513$\pm$0.009 & 0.514 \\ 
 1400 & 10$^8$-10$^9$ & 0.574$\pm$0.009 & 0.572 \\
 \hline
\end{tabular}
\caption{Intermediate scaling exponent $\beta_{pow}$ obtained through a power law fit $\Nu\sim\Ra^{\beta_{pow}}$ by \citet{Akhmedagaev2020} over the range given by $\Ra$ for each $\Ha$, compared to the average exponent predicted by the $\erfc$ model over the given range of $\Ra$ calculated as $\beta_{\erfc} = \langle \partial \log(\Nu)/\partial \log(Ra) \rangle$.}
\label{Table1}
\end{table}

As can be seen from the comparisons shown in the figures, the erfc-model captures both asymptotic cases as it must, and for the low-$\Pran$ data, it also captures the entire smooth transition region very well. 
The universality of this transition is notable, since the numerous datasets shown here, obtained from both numerical simulations and experiments, have varying aspect ratios and boundary conditions.
For the $\Pran = 8$ case, the small overshoot at the beginning of the buoyancy-dominated regime that is not captured by the erfc-model is well described through the perturbative extension of it, the erfc-$\tau$ model, indicated by the black dashed line for a constant $\tau$. A notable deviation between the model and data that occurs at low $\Ra$ and high $\Ha$ for low $\Pran$ is due to the occurrence of wall modes with very little to no bulk convection, as discussed in Ref.~\cite{Teimurazov2024}. 

\begin{figure}[h]
    \centering
    \includegraphics[width = 0.5\columnwidth]{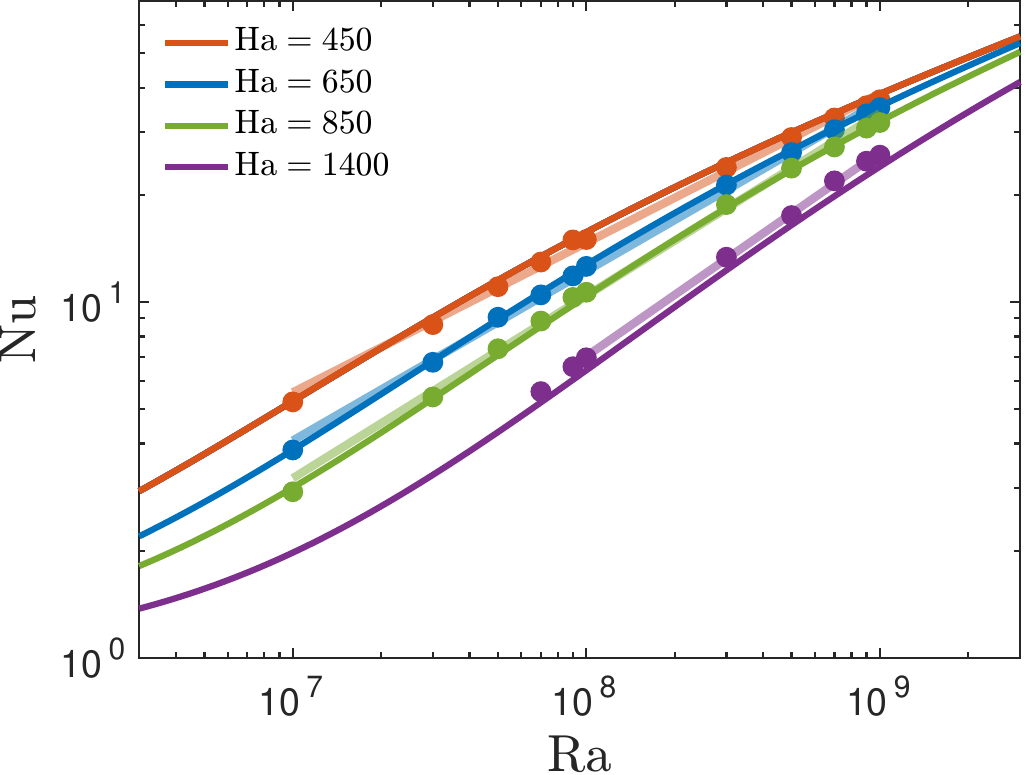}
    \caption{Comparison of the $\erfc$ model (solid lines) to the data (markers) and intermediate power-law fits (semi-transparent) of \citet{Akhmedagaev2020} plotted over their fitted range of $\Ra$. }  
    \label{fig:Akh-compare}
\end{figure}

For a stricter quantitative comparison between the model and data, we focus on the intermediate scaling regimes measured by Akhmedagaev et al. \cite{Akhmedagaev2020} for $10^7 \leqslant \Ra \leqslant 10^9$ and $450 \leqslant \Ha \leqslant 1400$ summarised in table \ref{Table1}.  In Fig.~\ref{fig:Akh-compare}, we show the results of the model (solid lines) in reproducing the data from Ref.~\cite{Akhmedagaev2020}. We additionally plot the measured intermediate scaling laws $\Nu \sim Ra^{\,\beta}$ of Ref.~\cite{Akhmedagaev2020} using the stated exponents and prefactors over the range of their fit. In table \ref{Table1}, we state the exponents explicitly and compare to the average predicted exponent of our model over the range of $\Ra$ fitted by in Ref.~\cite{Akhmedagaev2020}, calculated as $\beta_{\erfc} = \langle \partial \log(\Nu)/\partial \log(Ra) \rangle$, and see that our model reproduces these exponents within error bars in nearly all cases.

\begin{figure}[H]
    \centering
    \includegraphics[width = 0.5\columnwidth]{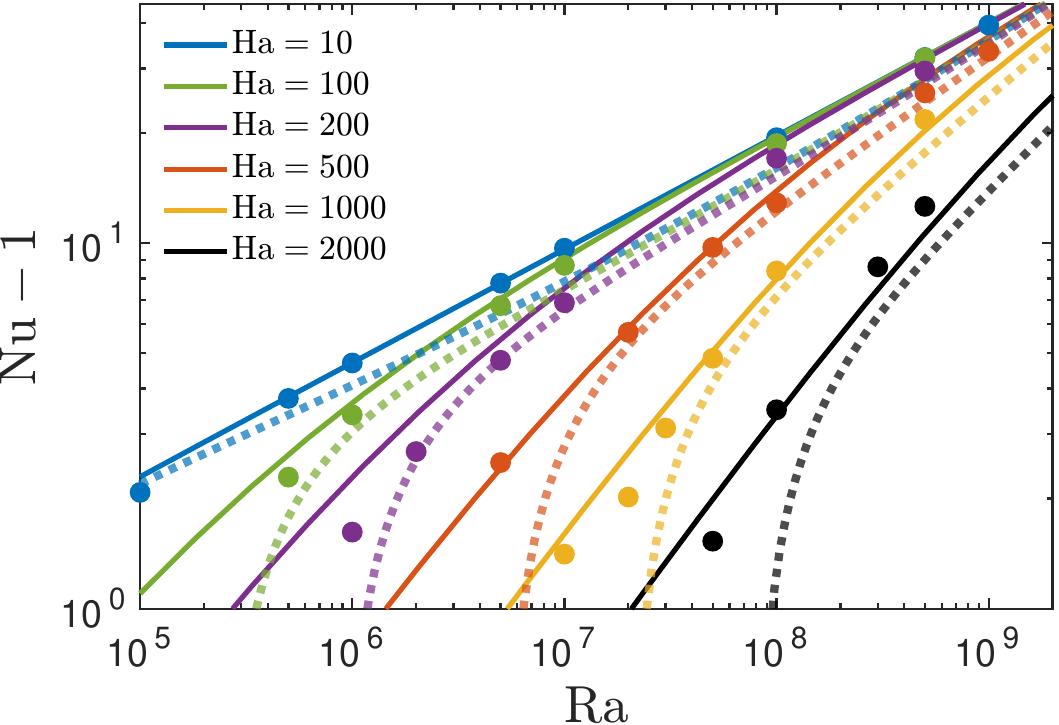}
    \caption{Comparison of the $\erfc$ model (solid lines) to the magnetoconvection extension of the Grossmann-Lohse model (dotted lines) from Z\"urner \cite{Zuerner2020b} and our DNS data (markers) \cite{Teimurazov2024}.}  
    \label{fig:GL_compare}
\end{figure}

In Fig.~\ref{fig:GL_compare} we present a comparison between our data, the erfc-model and the results of the Z\"urner \cite{Zuerner2016, Zuerner2020b} extension of the Grossmann-Lohse theory for $\Pran = 0.025$. As can be seen from the data shown in the figure, at higher values of $\Ha$ the results are reasonably similar although our model is typically closer to the data than the Z\"urner model. The Z\"urner model has a systematic deviation from the data at low to mid $\Ha$, particularly at higher values of $\Ra$.

\subsection{Full magnetoconvection}
\begin{figure}
    \centering
    \includegraphics[width = 0.6\columnwidth]{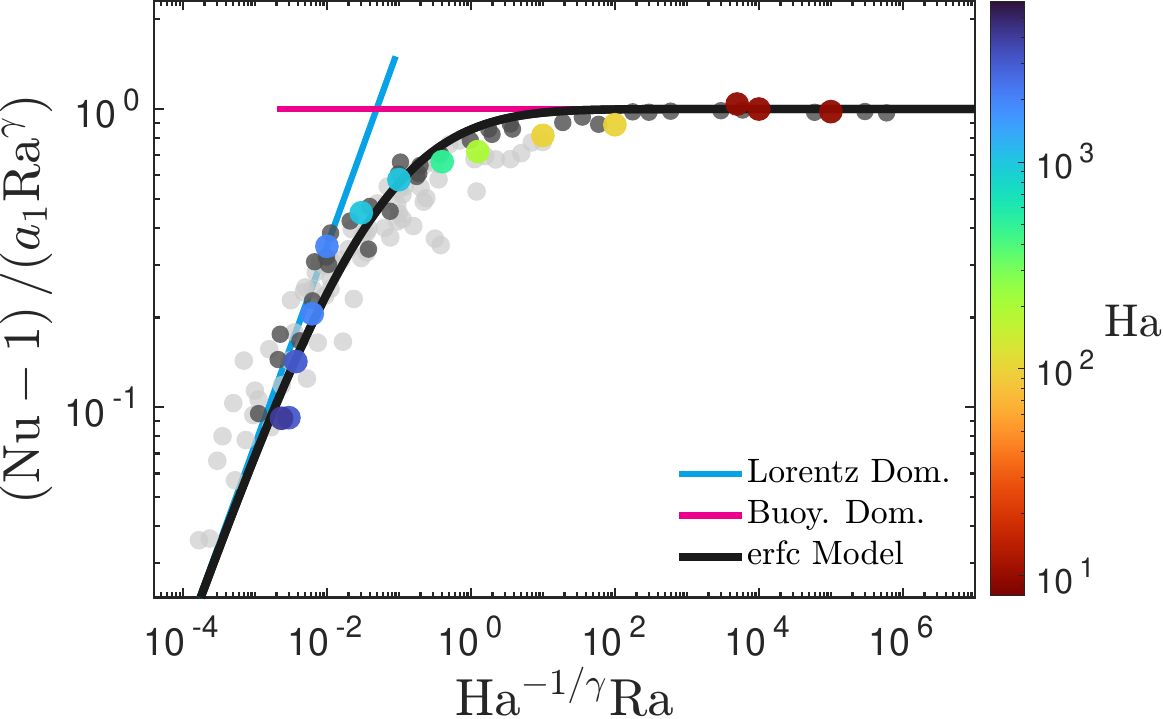}
    \caption{Comparison between model and data for full 3D MC simulations (coloured) for $10\leq\Ha\leq3000$, full 2D MC simulations (light grey) for $10^2\leq\Ha\leq 10^5$ at $\Pran = 0.25$, $\Pm = 0.5$ \cite{cresswell2023force}, and quasistatic 3D simulations for $10\leq\Ha\leq2000$ at $\Pran = 0.025$ (dark grey) \cite{Teimurazov2024}.  Note that the vertical axis has been compensated by the prefactor $a_1$ for direct comparison between the data sets.}  
    \label{fig:full_mhd}
\end{figure}

As the derivation of the model does not rely on the quasistatic approximation (\S \ref{sec:heat_transp_model}), it should be applicable to full MC with an adjustment of the prefactor $a_1$ only. 
In Fig.~\ref{fig:full_mhd} we show data obtained by simulations of the full magnetoconvection system in a periodic layer of unit aspect ratio at $\Pr=0.25, \Pm = 0.5$ in comparison with our model and data from the quasistatic regime. 
The parameters used in the $\erfc$ model here are shown in table \ref{tab:full_mhd_fit}.
We show the data points obtained by our DNS as coloured markers, two-dimensional simulations of various aspect ratios from Ref.~\cite{cresswell2023force} as light grey markers, and for comparison the $\Pr=0.025$ quasistatic simulation data from \cite{Teimurazov2024} (see also Fig.~\ref{fig:quasi}, top panel) in dark grey, the latter with appropriately adjusted $\gamma$ and prefactor $a_1$ to account for the difference in $\Pr$.
As can be seen from the comparison, the model captures the transition between the Lorentz-force-dominated and the buoyancy-dominated regimes also for magnetoconvection at finite magnetic Reynolds number where magnetic-field fluctuations have to be taken into account, suggestive of further universality in the transition.

\begin{table}[H]
\centering
\begin{tabular}{l c c c c c} 
 \hline
 \quad\:\: Model & $\gamma$ & $a_1$ & $a_2$ & $b$   \\ 
 \hline
 $\Pr=0.25$ ($\erfc$)    & 0.33 & 0.087 & 0.65 & 0.55   \\  
 \hline
\end{tabular}
\caption{Overview of the parameters used in the $\erfc$ model for full MC at the studied values of $\Pr$.  Parameters $\gamma$, $a_1$ and $a_2$ are known from existing literature.  The quantity $b$ is the free parameter of the theory.}
\label{tab:full_mhd_fit}
\end{table}

\section{Conclusions}

We have introduced a heat transport model with a single free parameter that quantitatively captures the transition between buoyancy-dominated and Lorentz-force-dominated regimes in magnetoconvection under a vertically aligned magnetic field.
The model provides a simple analytic expression which asymptotes to previously observed power laws in the buoyancy-dominated or Lorentz-force-dominated regimes and exhibits a smooth transition between these two extreme regimes.
The model has been validated through our own DNS, and existing numerical and experimental data in the literature in the quasistatic and full MC regimes. 
An extension of the model with an additional parameter describes the so-called overshooting effect observed at higher Prandtl numbers in the transitional region between the regimes of clear dominance of either gravitational buoyancy or the Lorentz force.

A notable feature of the model is that it suggests that the transition between the two extreme regimes of buoyancy and Lorentz-force dominance is smooth, and that previously observed discrete intermediate power laws may be viewed as the leading order contribution of the proposed model at a given local point in parameter space.  Thus, these power laws lie tangent to a smooth underlying surface in parameter space proposed by the model.

We finally mention that the proposed model may be straightforwardly applied to rotating convection and has the potential to be generalised to rotating magnetoconvection since the primary forces have a similar structure.
This task will be performed in our future studies.

\section*{Acknowledgements}
We thank G.~Vasil for helpful discussions and acknowledge financial support from the Deutsche Forschungsgemeinschaft (SPP1881 ``Turbulent Superstructures" and grants  Sh405/20, Sh405/22, Li3694/1). 
ML would like to thank the Isaac Newton Institute for Mathematical Sciences, Cambridge, for support and hospitality during the programme ``Anti-diffusive dynamics: from sub-cellular to astrophysical scales"  where work on this paper was undertaken. This work was supported by EPSRC grant no EP/R014604/1 and used the ARCHER2 UK National Supercomputing Service (https://www.archer2.ac.uk), with resources provided by the UK Turbulence Consortium (EPSRC grants EP/R029326/1 and EP/X035484/1).

\appendix

\section{Geometric interpretation of the collapse}
\label{sec:appendix_geo}
\begin{figure}[h]
    \centering
    \includegraphics[width = 0.75\columnwidth]{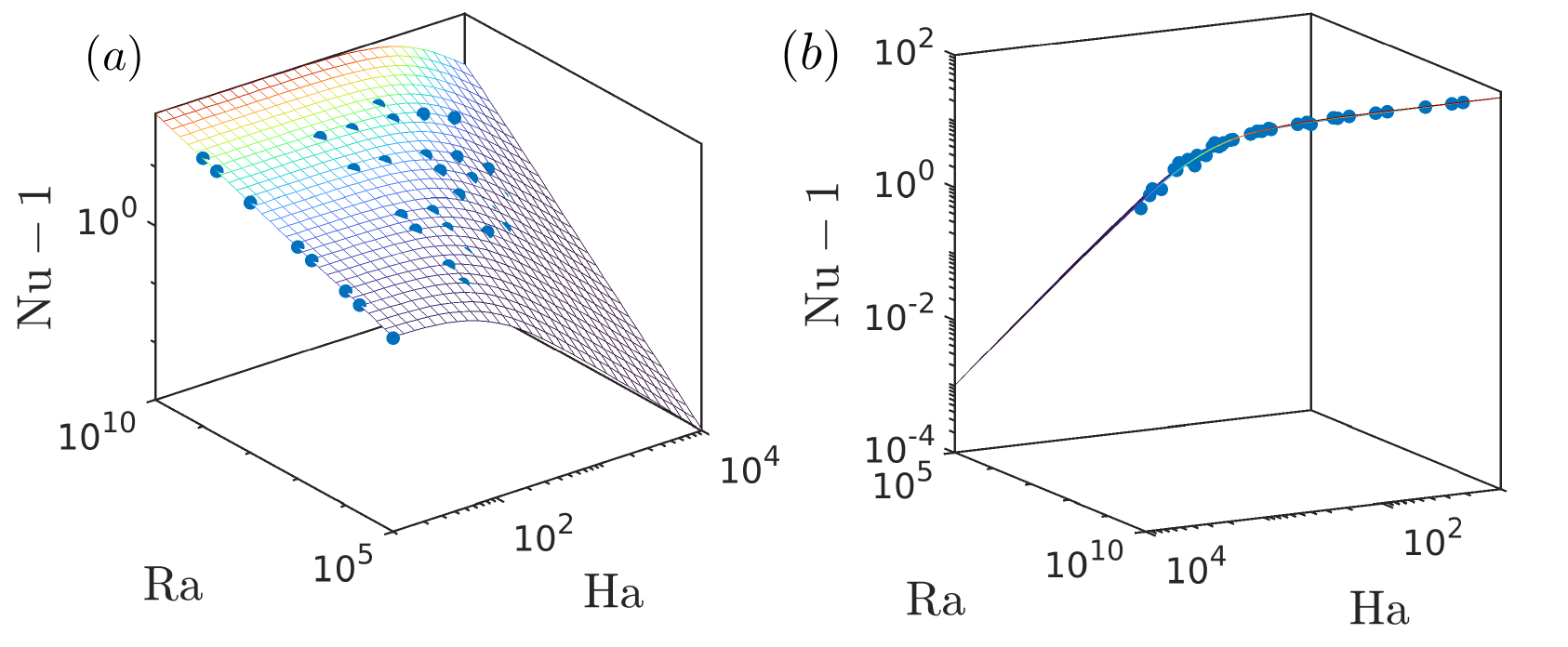}
    \caption{$(a)$ Typical view and $(b)$ rotated view of parametrised surface $\Nu = \Nu\,(\Ra,\Ha)$ with data from the quasistatic magnetoconvection simulations at $\Pran=0.025$ Ref.~\cite{Teimurazov2024}. The surface colour is denoted by the value of $\Nu$.  We note that from this particular rotation angle in $(b)$, the data is already collapsed under projection.} 
    \label{fig:rotated_collapse}
\end{figure}

The change of coordinates introduced in \S \ref{sec:transition-coordinates} i.e. $x=\log[\Ha^{-1/\gamma}\Ra]$ and $y=\log[(\Nu-1) \Ra^{-\gamma}]$ can be viewed as a linear transformation $\mathcal{T}$ on the three-dimensional parameter space $\bb{r} = (r,h,n) = (\log(\Ra),\log(\Ha),\log(\Nu-1))$ composed with a simple projection $\mathcal{P}$ onto the two-dimensional $\bb{x} = (x,y)$ subspace.  More precisely, the coordinate transform can be expressed as
\begin{equation}
    \begin{pmatrix}
        x \\ 0 \\ y
    \end{pmatrix}
    = 
    \mathcal{P}\,\mathcal{T}\bb{r} 
    = 
    \begin{pmatrix}
        1 & 0 & 0 \\
        0 & 0 & 0 \\
        0 & 0 & 1
    \end{pmatrix}
    \begin{pmatrix}
        1 & -1/\gamma & 0 \\
        0 & 1 & 0 \\
        -\gamma & 0 & 1
    \end{pmatrix}
    \begin{pmatrix}
        r \\ h \\ n
    \end{pmatrix}
    .
\end{equation}
Furthermore, the linear transformation $\mathcal{T}$ can be decomposed through an $RQ$ decomposition as $\mathcal{T}=RQ$ where $Q$ is an orthogonal matrix (an element of the group $\rm{SO}(3)$) and $R$ is upper triangular. Thus, $Q$ extracts the purely rotational part of the transformation $\mathcal{T}$, with $R$ being responsible for stretching and shearing of the space. Performing the decomposition gives
\begin{equation}
    Q = 
    \begin{pmatrix}
        z_1 z_2 & -z_1/\gamma & \gamma z_1 z_2 \\
        z_1 z_2/\gamma & z_1 & z_1 z_2 \\
        -\gamma z_2 & 0 & z_2
    \end{pmatrix}
\end{equation}
where $z_1 = (\gamma^{-2}+1)^{-1/2}$ and $z_2 = (\gamma^{2}+1)^{-1/2}$.  
We note that the rotation and projection without the additional stretching and shearing is largely responsible for the collapse we observe in the data sets and that even applying the operation $\mathcal{P}Q$ to the space is sufficient to collapse the data within experimental error. This can be observed clearly in the surface plots in Fig.~\ref{fig:rotated_collapse}.
Applying the operator $R$ stretches the abscissa but also crucially shears the axis that is linearly independent of the projection plane. This correction removes the small dependence of the abscissa on $\Nu$ and thus recovers the scaling relation which defines the coordinates, see Eq.~\eqref{eq:transition-scaling}.

\newpage

\section{Choice of sigmoid function} \label{sec:choice_of_sigmoid}
In \S\ref{sec:parameterisation-master-curve}, we chose to model the slope of the master curve $s(x)$ by the complementary error function.  This function was primarily chosen as it allowed the free parameters to be easily recovered in closed form.  However, other choices of sigmoid-shaped functions may equally be used to model the transition, and thus, the method provided defines a family of models.  These may conveniently be captured as the solutions to a differential equation.  We centre our coordinate system about $c$, defining $\mathrm{x} = x-c$.  Then, the family of model curves with these properties may be obtained through a nonlinear boundary value problem for the master curve slope $s(\rm{x})$
\begin{equation}
    \frac{d^2 s}{d\mathrm{x}^2} + f(\mathrm{x}) \frac{d s}{d\mathrm{x}} = 0,
    \label{eq:bvp-master-curve}
\end{equation}
where $f(\mathrm{x})$ is an odd, monotonic function with $f'(\mathrm{x})\geq 0$.  Equation \ref{eq:bvp-master-curve} is additionally equipped with boundary conditions $s(\mathrm{x}\rightarrow\infty)=0$, and $s(\mathrm{x}\rightarrow-\infty)=\xi-\gamma$.  Here, different choices of $f(\mathrm{x})$ correspond to different choices of sigmoid-shaped functions for the master curve slope $s(\mathrm{x})$.  For example $f(\mathrm{x}) \sim \mathrm{x}$ recovers a complementary error function, $f(\mathrm{x})\sim \mathrm{x}/(\mathrm{x}^2+1)$ recovers an inverse tangent function, or $f(\mathrm{x})\sim \tanh(\mathrm{x})$ recovers a hyperbolic tangent function.  Some choices of $f(\mathrm{x})$ involving the slope itself also produce sigmoids, for example, $f(\mathrm{x}) \sim [1 - 2s(\mathrm{x})]$ recovers the logistic function.  More general choices of $f(\mathrm{x})$ that have these properties result in non-standard sigmoid functions for $s(\mathrm{x})$, for example, the choice of $f(\mathrm{x})\sim \mathrm{x}^3$ results in a solution that can be expressed in terms of the incomplete Gamma function multiplied by a polynomial part.  We see that other examples, in this context, represent generalisations of the error function model since, assuming analyticity of $f(\mathrm{x})$, we may expand in terms of a Taylor series which to lowest order gives $f(\mathrm{x}) \sim \mathrm{x} + \cdots$ by imposing that $f(\mathrm{x})$ is odd and has a non-zero derivative at $\mathrm{x}=0$.  Thus, the error function model results from the simplest set of constraints on the differential equation to produce sigmoid-like models.
This, combined with the ease at which the free parameters may be recovered, makes the complementary error function the most natural choice in modelling the master curve.   Relaxing the constraint of the odd symmetry of $f(\mathrm{x})$, allows for master curves that are not symmetric about the $\mathrm{x}=0$ point.  For the choice of the error function model \emph{i.e.} $f(\mathrm{x})=\mathrm{x}$, we may also immediately deduce the power law behaviour from the differential equation in the buoyancy or Lorentz-force dominated regimes by noting that Eq. \ref{eq:bvp-master-curve} becomes \begin{equation}
    \frac{d s}{d\mathrm{x}} = 0,
\end{equation}
in the limit of $\mathrm{x}\rightarrow\pm\infty$, giving straight line solutions for the master curve, corresponding to power laws in log-space.
We finally note that inserting the master curve into equation \ref{eq:bvp-master-curve} as $s(\mathrm{x}) = y'(\mathrm{x})$ yields a differential equation for $\Nu$ in terms of $\Ra$ and $\Ha$ and thus, approximates local changes in the heat transport for given parameter values.

\section{Simulation data table} \label{sec:full_mhd_data}
\begin{table}[H]
\centering
\begin{tabular}{c c c c c c} 
 \hline
 $\Pran$ & $\Pm$ & $\Ha$ & $\Ra$ & $\Nu$ & $\Rm$ \\ 
 \hline
 0.25 & 0.5 & 10 & $5\times10^6$ & 16.45 & 826.5  \\ 
 0.25 & 0.5 & 10 & $1\times10^7$ & 19.74 & 988.9  \\
 0.25 & 0.5 & 10 & $1\times10^8$ & 40.56 & 2956.3  \\
 0.25 & 0.5 & 100 & $1\times10^7$ & 16.28 & 657.9 \\
 0.25 & 0.5 & 100 & $1\times10^8$ & 36.75 & 2218.6  \\
 0.25 & 0.5 & 200 & $1\times10^7$ & 14.49 & 507.6  \\
 0.25 & 0.5 & 500 & $5\times10^7$ & 22.33 & 935.9  \\
 0.25 & 0.5 & 1000 & $3\times10^7$ & 13.13 & 437.8  \\
 0.25 & 0.5 & 1000 & $1\times10^8$ & 24.40 & 995.4  \\
 0.25 & 0.5 & 2000 & $5\times10^7$ & 7.59 & 457.1  \\
 0.25 & 0.5 & 2000 & $8\times10^7$ & 13.98 & 601.7  \\
 0.25 & 0.5 & 3000 & $8\times10^7$ & 4.46 & 320.9  \\
 0.25 & 0.5 & 3000 & $1\times10^8$ & 6.75 & 415.4  \\
 \hline
\end{tabular}
\caption{Time-averaged global Nusselt number $\Nu$ and magnetic Reynolds number $\Rm = \Rey\Pm$, where $\Rey$ is the standard Reynolds number calculated with the root-mean-square velocity for each the direct numerical simulations in the full MC case.}
\label{tab:simuatlion_data}
\end{table}

\bibliographystyle{unsrt}

\end{document}